\documentclass[12pt]{article}
\usepackage[T1]{fontenc}

\usepackage{amsmath}
\usepackage{amssymb}
\usepackage{graphicx}
\usepackage{float} 
\restylefloat{table}
\usepackage{dcolumn}
\usepackage{bm}
\usepackage{CJKutf8} 
\usepackage[
backend=biber,
style=phys,%
    articletitle=false,biblabel=brackets,%
    chaptertitle=false,pageranges=false%
]{biblatex}

\usepackage[a4paper, total={6.5in, 10in}]{geometry}

\begin{document}

\newcommand{\la}{\langle}
\newcommand{\ra}{\rangle}
\newcommand{\bra}[1]{\langle{#1}\vert}
\newcommand{\ket}[1]{\vert{#1}\rangle}
\newcommand{\braket}[2]{\langle{#1}\vert{#2}\rangle}
\renewcommand{\vec}[1]{{\bm{#1}}}

\begin{center}
{\Large \textbf{Kekul\'e Moir\'e Superlattices}}\\

\medskip

Yusen Ye\textsuperscript{1},
Jimin Qian\textsuperscript{1},
Xiao-Wei Zhang\textsuperscript{1},
Chong Wang\textsuperscript{1},
Di Xiao\textsuperscript{1,2*},
Ting Cao\textsuperscript{1*}.
\\
\bigskip
\textsuperscript{1} Department of Materials Science and Engineering, University of Washington, Seattle, WA, USA
\\
\textsuperscript{2} Department of Physics, University of Washington, Seattle, WA, USA
\\
\bigskip

* Corresponding Authors. Emails: dixiao@uw.edu, tingcao@uw.edu.

\end{center}

\bigskip

\begin{abstract}
Moir\'e superlattices from stacks of van der Waals materials offer an exciting arena in the fields of condensed matter physics and materials science.
Typically, these moir\'e superlattices consist of materials with identical or similar structures, and the long moir\'e period arises from a small twist angle or lattice mismatch. 
In this article, we discuss that long moir\'e period appears in a new moir\'e system by stacking two dissimilar van der Waals layers with large lattice mismatch, resulting in couplings between moir\'{e} bands from remote valleys in the momentum space. 
In this system, the first layer is reconstructed using a $\sqrt{3}$ by $\sqrt{3}$ supercell that resembles the Kekul\'{e} distortion in graphene, and such reconstruction becomes nearly commensurate with the second layer. 
This Kekul\'e moir\'e superlattice is realized in heterostructures of transition metal dichalcogenides and metal phosphorus trichalcogenides such as MoTe$_2$/MnPSe$_3$. 
By first-principles calculations, we demonstrate that the antiferromagnetic MnPSe$_3$ strongly couples the otherwise degenerate Kramers' valleys of MoTe$_2$, resulting in valley pseudospin textures that depend on N\'eel vector direction, stacking geometry, and external fields. 
With one hole per moir\'e supercell, we predict that the system can become a Chern insulator, of which the topology is tunable by external fields.
\end{abstract}

\bigskip

\section{Introduction}\label{sec1}
Moir\'e superlattices are usually formed by stacking van der Waals (vdW) materials with a small twist angle and/or a small lattice mismatch. 
The nearby electronic states in the momentum $\vec{k}$-space are coupled by the long moir\'e period and, in the case of graphene and transition metal dichalcogenides, resulting in two independent copies of moir\'e bands from each valley \cite{bistritzer2011moire,shallcross2010electronic,wu2019topological,zhang2021spin,wu2018hubbard}. 
These moir\'e bands, combined with spin degrees of freedom as in a TMD moir\'{e}  bilayers, can bring about a variety of correlated and topological phases\cite{tang2020simulation,cao2018correlated,cao2018unconventional,li2021quantum,wang2022one,hunt2013massive,regan2020mott,seyler2019signatures}.
However, the short-range nature of the moir\'{e} scattering potential, together with the valley-spin locking effect, prevents direct coupling between the two Kramers' valleys (i.e., $\ket{\vec{K},\uparrow}$ and $\ket{\vec{K}',\downarrow}$) from the same layer\cite{xu2014spin}.
Since the moir\'e effects modulate these valleys independently, it remains challenging to fully control and utilize the Kramers' valley-spin pairs within the moir\'{e} platform.\\
\indent Prior attempts to connect remote valleys have focused on correlation effects and lattice reconstruction. 
On the one hand, the correlation effects have the potential to induce a coherent superposition between valley states, but controlling the strength and the form of correlation is challenging for the Kramers' valley-spin pairs. 
On the other hand, direct intervalley coupling can be accomplished by $\vec{k}-$space scattering on the scale of the large reciprocal lattice vector in a structure without moir\'{e} effects.
For example, in Kekul\'{e} distorted graphene, the $\sqrt{3}\times\sqrt{3}$ reconstruction of the unit cell generates scattering potential that connects $\vec{K}$ and $\vec{K}'$, giving rise to intervalley coherent states \cite{gutierrez2016imaging,ryu2009masses,hou2007electron,liu2014intervalley}. 
Kekul\'e graphene has been achieved experimentally through the use of adatoms to form an effective $\sqrt{3}\times\sqrt{3}$ lattice\cite{gutierrez2016imaging,bao2021experimental}, or by stabilizing a Kekul\'{e} bond-ordered graphene through interaction when Landau levels appear under high magnetic field\cite{liu2022visualizing,bultinck2020ground}. 
However, as the Kekul\'{e} reconstruction in graphene relies on structure modifications and electron correlations, rational control of the intervalley scattering at the moir\'e periodicity remain challenging. 
It is also a non-trivial task to extend the Kekul\'{e} effects to other moir\'e systems and establish a tunable coupling in a Kramers' valley-spin pair to navigate the fascinating moir\'e space. \\
\indent In this article, we discuss a new type of moir\'{e} system that goes beyond the limits of small twist angle and small lattice mismatch, and, in doing so, short-range and long-range scatterings in $\vec{k}-$space emerge spontaneously and simultaneously. 
By first-principles calculations and continuum modeling, we show that the coupling between Kramers' valley-spin states induces a pseudospin topological texture at the moir\'{e} scale in MoTe$_2$/MnPSe$_3$ superlattices.
We show that the coupling and the topological texture can be fully controlled by magnetic order and external fields, leading to topologically non-trivial bands and a quantum anomalous Hall insulator phase at one hole per moir\'e cell.\\

\section{Results and Discussion}\label{sec2}
\subsection{Kekul\'e Moir\'e Structure and Intervalley Coupling}
\begin{figure}[H]
\includegraphics[width=\columnwidth]{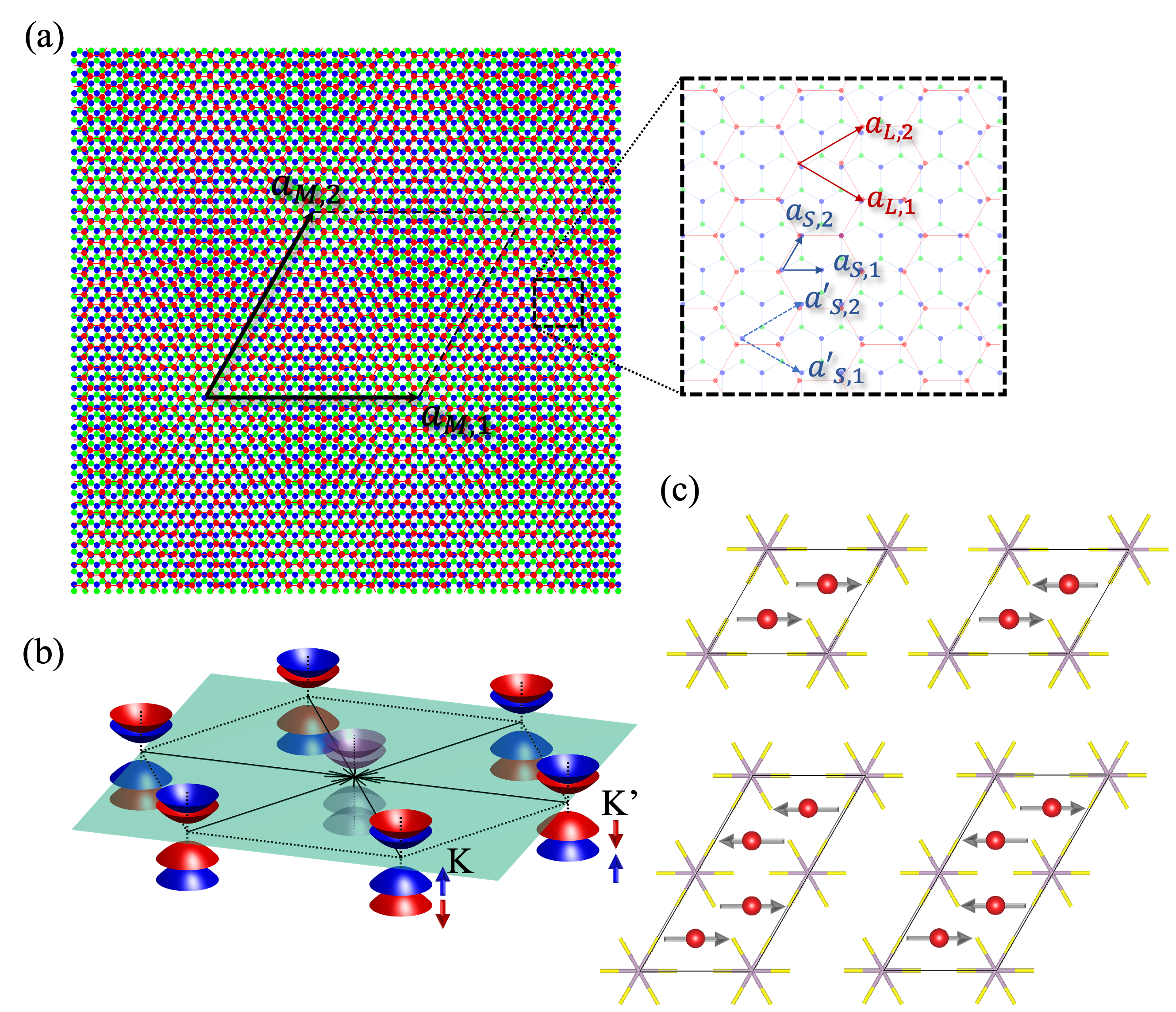}
\caption{\label{fig:LatticeStruct} (a) The moir\'{e} pattern formed by stacking MPX$_3$ on top of TMD with a $30^\circ$ twist angle. Blue (red) circles represent the metal ions in TMD (MPX$_3$). Inset: $\{\vec{a}_\text{S}\}$ ($\{\vec{a}_\text{L}\}$) denote the primitive lattice vectors of TMD (MPX$_3$), where $\sqrt{3}\|\vec{a}_\text{S}\|\approx \|\vec{a}_\text{L}\| $ as shown in dashed $\{\vec{a}'_\text{s}\}$. (b) A schematic plot of monolayer TMD band structure showing the valleys. The $\vec{K}$ ($\vec{K}'$) valleys of monolayer TMD are locked with spin up (down) as indicated by blue (red) coloring. Under Kekul\'{e} reconstruction, both valleys are folded back to $\vec{\Gamma}$. (c) Possible magnetic orders of MPX$_3$ are shown. Arrows denote local magnetic moments. From upper left going clockwise: FM, AFM N\'{e}el, AFM stripy, AFM zigzag.}
\end{figure}

\indent The new moir\'{e} superlattice consists of two distinct layers of two-dimensional (2D) hexagonal lattices. The ratio between their lattice constants, $a_\text{L}/a_\text{S}$, is about $\sqrt{3}$. We then reconstruct a $\sqrt{3}a_\text{S} \times \sqrt{3} a_\text{S}$ Kekul\'e cell of the smaller lattice, such that the two layers become almost commensurate as shown in Fig. \ref{fig:LatticeStruct}(a). 
We redefine the lattice mismatch to be measured between $\sqrt{3}a_\text{S}$ and $a_\text{L}$, and the twist angle to be measured starting from $30^\circ$. 
This construction, which we coin the name as Kekul\'e moir\'e superlattices, can be achieved by combining any pair of layered vdW materials equipped with a hexagonal Bravais lattice, as long as their lattice constant ratio is about $\sqrt{3}$. 
As this work aims to couple and control the Kramers' valley-spin pairs, we choose the first layer with $a_\text{S}$ to be TMD, which has been known for their valley-spin degrees of freedom\cite{xiao2012coupled,cao2012valley}. 
However, even with a spin-conserving Umklapp scattering from the lattice potential of the second layer, which folds both $\vec{K}_\text{TMD}$ and $\vec{K'}_\text{TMD}$ back to $\vec{\Gamma}$, the spin-valley locking effect will forbid any coupling between $\ket{\vec{K},\uparrow}$ and $\ket{\vec{K}',\downarrow}$.  
As a result, the heterostructure will not develop any intervalley coupling if a spin-flipping mechanism, e.g., magnetism or spin-orbit coupling, is not involved. 
Fortunately, a class of magnetic vdW materials, metal phosphorus trichalcogenides (MPX$_3$), has lattice constants approximately matching $\sqrt{3}a_\text{TMD}$, and is therefore used as the second layer. 
As an important class of vdW materials, MPX$_3$ has rich magnetic orders, including ferromagnetic (FM), zigzag anti-ferromagnetic (AFM), and N\'{e}el AFM depending on the atomic species of M and X\cite{sivadas2015magnetic,chittari2016electronic}. 
Based on optimal lattice matching conditions, several combinations of TMD and MPX$_3$ that can give rise to long moir\'e period are listed in Table \ref{tab:MPX3/TMD}.\\ 

\indent The effective potential of MPX$_3$ that acts on TMD can be expanded as 
\begin{equation*}
V(\vec{r})=\sum_{\vec{G}_\text{L}} e^{i\vec{G}_\text{L}\cdot \vec{r}}\vec{V}_{\vec{G}_\text{L}}\cdot\vec{\sigma},
\end{equation*}
where $\vec{G}_\text{L}$ is a reciprocal lattice vector of MPX$_3$, $\vec{V}_{\vec{G}_\text{L}}\equiv(V_{0,\vec{G}_\text{L}},V_{x,\vec{G}_\text{L}},V_{y,\vec{G}_\text{L}},V_{z,\vec{G}_\text{L}})$ is the Fourier transformed lattice potential in a four-vector form to denote the spin-independent and spin-dependent components, and $\vec{\sigma}\equiv(\sigma_0,\sigma_x,\sigma_y,\sigma_z)$ is a four vector formed by $\mathbb{I}_2$ and Pauli matrices which act on the spinor Hilbert space. 
To this end, many previous works have studied the proximity effects of a magnetically ordered vdW material layered on TMD, but these efforts have mostly used the simple analog of homogeneous magnetic field, i.e., $\vec{V}_{\vec{G}_\text{L}}\cdot\vec{\sigma}$ at $\vec{G}_\text{L} = 0$\cite{seyler2018valley,onga2020antiferromagnet,zhang2020abundant,zhao2017enhanced,scharf2017magnetic}. 
In Kekul\'e moir\'e superlattice, to couple the two valleys in both momentum and spinor space, we need a lattice scattering momentum $\vec{G}_\text{L}=\pm(\vec{K}-\vec{K'})$, as well as a spin-flipping matrix $\sigma_x$ or $\sigma_y$. 
Combined together, the condition is $V_{x/y,\pm (\vec{K}-\vec{K'})}\neq0$. 
This requires an in-plane magnetic order in MPX$_3$, as a sizeable $V_{x,\vec{G}_\text{L}}$ or $V_{y,\vec{G}_\text{L}}$ can be obtained from Fourier transforming an effective exchange field that originates from spin polarization aligned in-plane. Further, a more detailed symmetry analysis (see appendix) shows that only FM and N\'eel ordered AFM can generate an intervalley coupling.
Based on this phenomenological theory, we select MoTe$_2$ and MnPSe$_3$ (in-plane AFM ordered) from Table \ref{tab:MPX3/TMD} to illustrated the Kekul\'e moir\'e physics in a realistic heterostructure.\\

\begin{table}[H]
\begin{center}
\begin{minipage}{\textwidth}
\caption{Lattice constant, ground state magnetic order (MO) and best lattice-matching TMDs. (Lattice constants are obtained from DFT) }\label{tab:MPX3/TMD} 
\begin{footnotesize}
\begin{tabular*}{\textwidth}{|@{\extracolsep{\fill}}c|c|c|c|c|c|c@{\extracolsep{\fill}}|}
\hline 
 & MnPS$_3$& MnPSe$_3$& FePS$_3$
&CoPS$_3$ &NiPS$_3$&NiPSe$_3$\\ \hline \hline
a(\AA)& $6.08$ & $6.38$ & $5.94$ & $5.91$ & $5.82$ & $6.13$\\ \hline
MO\cite{ouvrard1985structural,joy1992magnetism,le1982magnetic,wildes2017magnetic} & N\'{e}el(z) & N\'{e}el(xy) & zigzag(z) & zigzag(x) & zigzag(x) & zigzag(x)\\\hline
TMD match& MoTe$_2$& MoTe$_2$ & MoTe$_2$, WSe$_2$ & MoSe$_2$, WSe$_2$ & MoSe$_2$, WSe$_2$ & MoTe$_2$ \\
\hline
\end{tabular*}
\end{footnotesize}
\end{minipage}
\end{center}
\end{table}

\subsection{MoTe$_2$/MnPSe$_3$ Heterostructure}
In the MoTe$_2$/MnPSe$_3$ heterostructure, the lattice mismatch is about $ 4 \%$. 
The resulting moir\'{e} wavelength can be as large as $8$ nm, given by $a_\text{MoTe$_2$}/\delta$, where $\displaystyle\delta=\frac{\vert a_\text{MnPSe$_3$}-\sqrt{3} a_\text{MoTe$_2$}\vert}{a_\text{MnPSe$_3$}}$ (see appendix). 
The moir\'{e} lattice vectors align with the primitive lattice vectors of MoTe$_2$ as illustrated schematically in Fig.\ref{fig:LatticeStruct}(a). 
Similar to TMD bilayers that can be either R-type or H-type, the Kekul\'e moir\'e heterostructure can also be constructed from either $0^\circ$ or $60^\circ$ twist angle (measured from $30^\circ$). 
The two constructions are the same up to a total rotation if either layer is inversion symmetric; if both layers break inversion symmetry, the $0^\circ$ and $60^\circ$ constructions are of different structures and thus will host different electronic behaviors in general. 
Since the lattice of MnPSe$_3$ contains inversion symmetry, the  $0^\circ$ and $60^\circ$ constructions are equivalent in the MoTe$_2$/MnPSe$_3$ heterostructure.\\
\indent To study the MoTe$_2$/MnPSe$_3$ Kekul\'e moir\'e superlattice that consists of over $3000$ atoms in one moir\'e supercell, we performed first-principles calculations (see appendix) on $12\times 12$ commensurate structures based on local stacking geometry, followed by continuum modeling using parameters from the calculations. 
Among these commensurate unit cells are three high-symmetry stackings shown in Fig. 
 \ref{fig:MoireDia}(c).
The DFT-PBE band structure of stacking A shows a type-II band alignment in Fig.\ref{fig:MoireDia}(b). 
At $\vec{\Gamma}$, the band edge states are formed by MoTe$_2$, and the $\vec{K}$ and $\vec{K}'$ valleys are folded to the center of the Kekul\'e Brillouin zone by the Umklapp scattering. 
As the valence band top from MoTe$_2$ is well separated from the MnPSe$_3$ bands located at $0.66$ eV below, the two topmost valence bands become ideal candidates to analyze and control intervalley coupling in the Kekul\'e moir\'e superlattice.
The band alignments in other commensurate stacking geometries are generally similar to Fig.\ref{fig:MoireDia}(c) except for band splittings and energy variations on the order of tens of meV, which arise from the intervalley coupling and intravalley potential that we will discuss next. \\

\subsection{Moir\'e Hamiltonian and Pseudospin Texture}
\begin{figure}[H]
\includegraphics[width=\columnwidth]{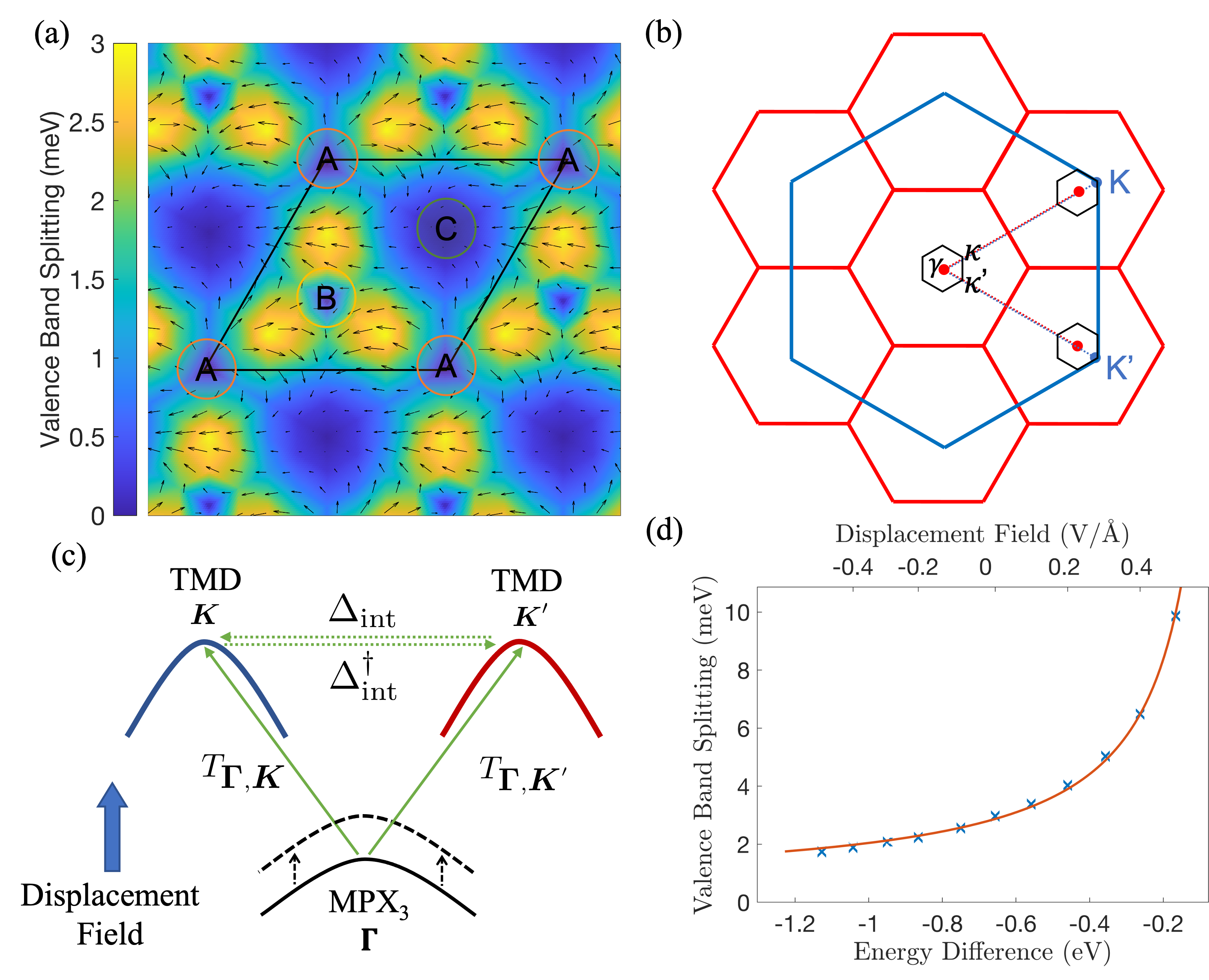}
\caption{\label{fig:MoireOffD}(a) Intervalley coupling induced splitting between MoTe$_2$ valence bands in the real-space MoTe$_2$/MnPSe$_3$ moir\'{e} superlattice from first-principles calculation on a $12 \times 12$ grid. The deep blue color shows no intervalley splitting while the yellow color shows largest intervalley splitting up to $3.0$ meV. Arrows represent the strength and the direction of effective valley pseudospin magnetic field from model as defined in the main text. (b) The formation of moir\'{e} Brillouin zone. The reciprocal lattice vectors of MPX$_3$ (red) fold the $\vec{K}$/$\vec{K}'$ points back around $\vec{\Gamma}$ point as $
\vec{\kappa}$/$\vec{\kappa}'$ points. The resulting moir\'{e} Brillouin zone is parallel to the TMD (blue) Brillouin zone. (c) Schematic physical origin of the intervalley coupling (green dashed line) mediated by the magnetic substrate, through a two-step hopping (green solid lines) between $\ket{\vec{K}}$ and $\ket{\vec{K}'}$ valence valleys of TMD. Dashed parabolas represent bands being pushed up by an out-of-plane displacement field shown as the large blue arrow, which enlarges the intervalley coupling. (d) Valley splitting inversely proportionate to the energy difference between $\ket{\vec{K},\text{MoTe}_2}$ and $\ket{\vec{\Gamma},\text{MnPSe}_3}$ bands, where the energy difference is tuned by the out-of-plane displacement field as shown in the top $x$-axis. Crosses and the line are from first-principles results and fitting, respectively. In the fitting, we find $y=-\frac{1.58}{x}+0.46$, where $y$ is the valence band splitting, and $x$ is the energy difference between relevant bands from MoTe$_2$ and MnPSe$_3$.} 
\end{figure}

In the Kekul\'e moir\'e superlattice, the valley-spin polarized Bloch states of MoTe$_2$, i.e., $\{\ket{\vec{K}, \uparrow},\ket{\vec{K}',\downarrow}\}$, are mapped to $\pm\vec{\kappa}$ in the moir\'e Brillouin zone (Fig.\ref{fig:MoireOffD}(b)). We thus use an effective moir\'e Hamiltonian in the basis of $\ket{\vec{k}\mp\vec{\kappa}}$ to describe the low-energy physics of the moir\'e valence bands:
\begin{align}\label{eqn:Ham}
    H_\text{eff}(\vec{r})=\begin{bmatrix}
    H_{\vec{k}-\vec{\kappa}}+V_\text{d}(\vec{r}) & \Delta_\text{int}(\vec{r}) \\
    \Delta_\text{int}^\dagger(\vec{r}) & H_{\vec{k}+\vec{\kappa}}+V_\text{d}(\vec{r})
    \end{bmatrix}.
\end{align}

\begin{figure}[H]
\includegraphics[width=\columnwidth]{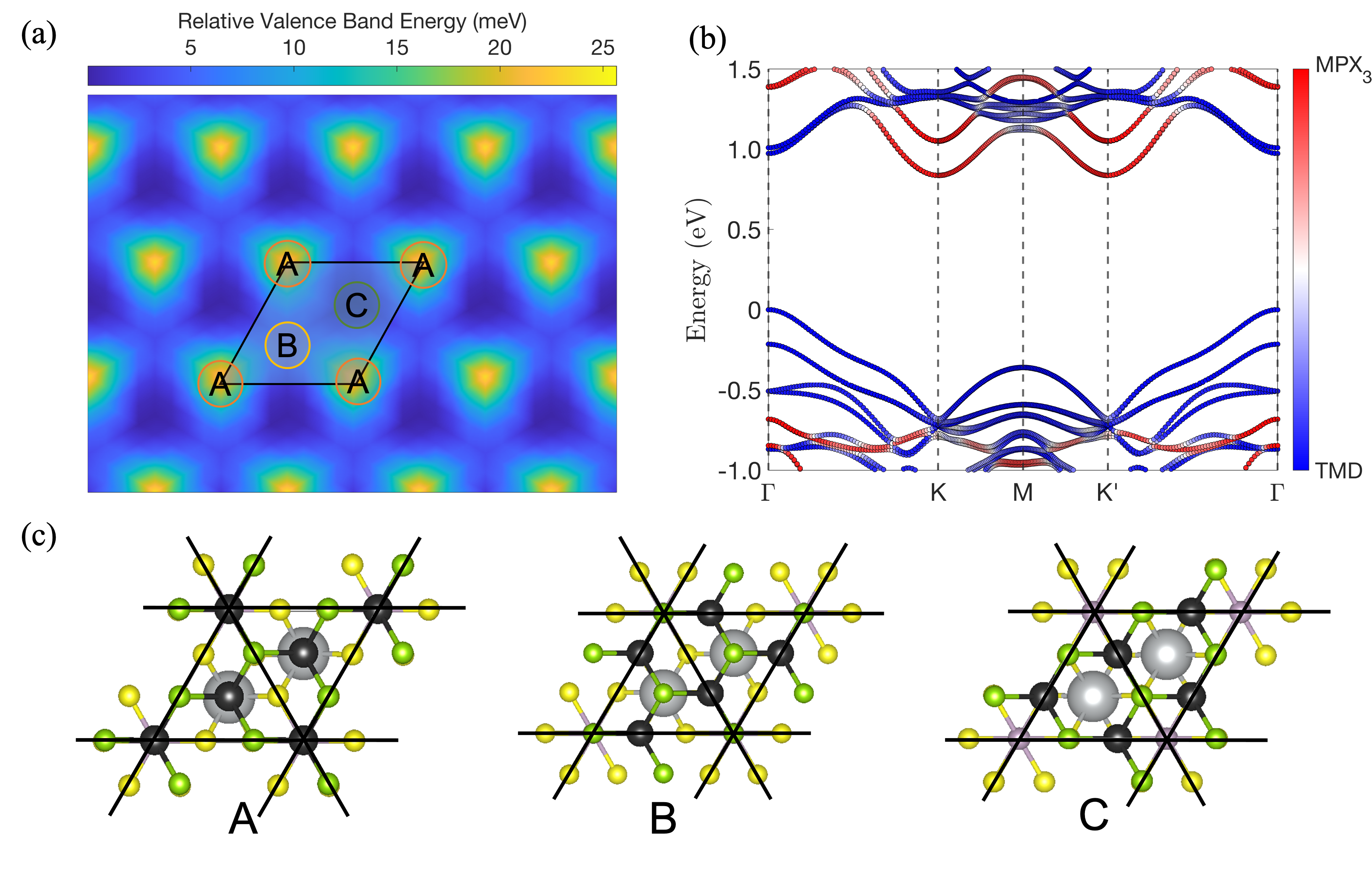}
\caption{\label{fig:MoireDia} (a) Kohn-Sham DFT valence band edge energies in the MoTe$_2$/MnPSe$_3$ moir\'{e} superlattice. A,B,C are high symmetry stacking regions that are shown in (c). (b) Kohn-Sham DFT bandstructure of MoTe$_2$/MnPSe$_3$ at high-symmetry stacking A. Blue (red) shows the projection of bands to the MoTe$_2$ (MnPSe$_3$) orbitals. (c) Three high symmetry stackings of TMD/MPX$_3$ heterostructure in top view. There exists mirror symmetries, $\sigma_\text{V}$, in every high symmetry local structure illustrated by the black solid lines. Black and green spheres illustrate transition metal and chalcogen atoms in the TMD layer, respectively; silver, yellow, and light purple spheres illustrate transition metal, chalcogen, and phosphorus atoms in MPX$_3$ layer, respectively.}
\end{figure}

We use the effective mass approximation: 
$H_{\vec{k}\mp\vec{\kappa}}=-\frac{\hbar^2(\vec{k}\mp\vec{\kappa})^2}{2m^*}$, where $m^*=0.86m_e$ from first-principles calculations. 
$V_\text{d}(\vec{r})$ depicts the intravalley moir\'e potential, which reads\cite{wu2018hubbard,wu2019topological}:
\begin{align}
    V_\text{d}(\vec{r})=2V\sum_{j=1,3,5}\cos(\vec{g}_{M,j}\cdot\vec{r}+\psi),
\end{align}
where $V$ is the structural potential amplitude, $\vec{g}_{M,j}$ are $(j-1)\pi/3$ counter-clockwise rotated from $\frac{4\pi}{\sqrt{3}L_M}\hat{y}$, and $\psi$ is a phase factor parameter. 
$V$ and $\psi$ can be obtained by fitting to the energy of valence band maxima in the $12 \times 12$ commensurate stackings, using the first-principles results in Fig.\ref{fig:MoireDia}(a). 
We find $V=2.3$ meV and $\psi=-16.6^\circ$. \\
\indent $\Delta_\text{int}(\vec{r})$ describes the coupling between the Kramers' valleys in the same layer, which is a virtual process that would have been absent without the Kekul\'e moir\'e effects. We note that similar forms of the Hamiltonian have appeared in previous studies in TMD moir\'e structures, but the off-diagonal terms represent a real scattering process from a valley in one layer to an adjacent valley in the other layer\cite{wang2017interlayer,zhang2021spin,wu2019topological}. The amplitude of $\Delta_\text{int}$ can be obtained directly from the first-principles calculations, which equals to half of the splitting between the top two valence bands in commensurate heterostructures shown in Fig.\ref{fig:MoireOffD}(a).
We find maximum band splitting of $3.0$ meV in the region between A and B, but the splitting vanishes at the three high-symmetry stackings.
To understand this pattern, we notice that all the three high-symmetry structures contain lattice mirror symmetries as shown in Fig.\ref{fig:MoireDia}(c). 
In the cases that the N\'eel vectors are in the mirror planes, degeneracy between the $\ket{\vec{K}  \uparrow}$ and $\ket{\vec{K'} \downarrow}$ valleys is protected. 
If the N\'eel vector rotates away, the lattice mirror symmetries still forbid intervalley coupling in the orbital parts of wavefunction, as long as the spin-orbit term in the interlayer coupling Hamiltonian is weak. 
This analysis has also been confirmed by our DFT calculations. \\
\indent Equation (\ref{eqn:Ham}) suggests that $\Delta_\text{int}$ can be recognized as a pseudospin magnetic field acting on the valley pseudospin as $\Delta_\text{int}=\Re(\Delta_\text{int})\tau_x-\Im(\Delta_\text{int})\tau_y$, where $\tau$'s are the Pauli matrices. 
As a result, $\Delta_\text{int}$ not only carries an amplitude corresponding to the size of the pseudospin magnetic field, but also a phase related to the direction of this field in the 2D plane.
Therefore, highly interesting valley pseudospin textures has appeared in the moir\'e supercell as shown in Fig.\ref{fig:MoireOffD}(a).\\
\indent In MoTe$_2$/MnPSe$_3$ Kekul\'e moir\'e superlattices, $\Delta_\text{int}(\vec{r})$ can be approximated in the first few harmonic expansions by noticing the translational symmetry and $\hat{\mathcal{C}_3}$ rotational symmetry\cite{zhang2021spin,wu2019topological}. 
Furthermore, as $\Delta_\text{int}(\vec{r})$ vanishes at the high symmetry stackings, the approximate form requires at least the first two harmonic expansions, which reads

\begin{equation}
    \begin{aligned}\label{eqn:OffDiaPot}
        \Delta_\text{int}(\vec{r})&=\Delta(1+\omega e^{-i(\vec{g}_{M,1}\cdot \vec{r})}+\omega^2 e^{-i\vec{g}_{M,2}\cdot \vec{r}})\\
        &-\Delta(e^{-i(\vec{g}_{M ,1}+\vec{g}_{M,2})\cdot \vec{r}}+\omega e^{-i(\vec{g}_{M,3}\cdot \vec{r})}+\omega^2 e^{-i\vec{g}_{M,6}\cdot \vec{r}}).
    \end{aligned}
\end{equation}
Here, $\omega=\braket{\vec{K},\uparrow}{\hat{\mathcal{C}}_3^{-1}(\vec{K},\uparrow)}\braket{\vec{K}',\downarrow}{\hat{\mathcal{C}}_3(\vec{K}',\downarrow)}=e^{i4\pi/3}$. 
The phase factor is obtained from both the orbital and the spinor part of the valley states\cite{zhang2021spin}. 
As $\ket{\pm \vec{K}}$ mainly consists of $\ket{d_{x^2-y^2}}\pm i\ket{d_{xy}}$ orbitals from the Mo ions, applying the $\hat{\mathcal{C}}_3$ results in an orbital phase factor of $\exp{\left(\pm i2\pi/3\right)}$. 
Besides, the $ \ket{\pm \vec{K}}$ valleys are spin-up(down) from the valley-spin locking, which generates an extra factor of $\exp{\left(\mp i\pi/3\right)}$. The effect of $\hat{\mathcal{C}}_3$ on the Bloch phases from $\ket{\pm \vec{K}}$ are canceled out by the same but opposite phase from the momentum carried by $\Delta_\text{int}$.
These together produce the overall phase of $\omega=e^{i 4\pi/3}$. Fitting equation (\ref{eqn:OffDiaPot}) to DFT calculations, we find $\Delta=-0.35$ meV. \\
\indent The valley pseudospin textures of a Kekul\'e moir\'e superlattice lead to winding patterns of spin polarization in the moir\'e frontier orbitals. At the conduction band edge where $\ket{\vec{K}, \uparrow}$ and $\ket{\vec{K}',\downarrow}$ share the same orbital wavefunction of $\ket{d_{z^2}}$ but differ in their spins, $\Delta_\text{int}$ represents a real-space magnetic field that dictates noncollinear spin orientations of Kekul\'e cells. At the valence band edge where the two valleys carry different orbitals and spins, $\Delta_\text{int}$ represents a set of magnetic field acting on each orbital-spin basis. As such the frontier orbital carries entangled orbitals and spins at the moir\'e periodicity.\\

\subsection{Controlling Pseudospin and Band Topology}
The Kekul\'e moir\'e Hamiltonian and the valley pseudospin textures can be controlled by electric field and magnetic order. 
In our calculations, we apply an out-of-plane displacement field to the MoTe$_2$/MnPSe$_3$ heterostructure with $0.1$ eV/\text{\AA} increment.
The amplitude of $\Delta_\text{int}$ (and effectively $\Delta$ as from equation (\ref{eqn:OffDiaPot})) increases dramatically under larger electric field pointing from the MPX$_3$ to TMD.
At the stacking region with maximum valley splitting, $\Delta_\text{int}$ increases from $3.0$ meV at no electric field to $5.0$ meV at $ + 0.3$ V/{\AA} electric field (Fig.\ref{fig:MoireOffD}(d)). 
To understand why electric field strongly couples to the pseudospin magnetic field strength, we analyze the microscopic interaction that leads to $\Delta_\text{int}$. 
In vdW heterostructures, electrons can hop between layers while conserving the in-plane $\vec{k}$. 
In TMD/MPX$_3$ Kekul\'e moir\'e, $\ket{\pm \vec{K},\text{TMD}}$ directly couples to $\ket{\vec{\Gamma}, \text{MPX}_3}$. 
We denote this hopping process as $T_{\vec{\Gamma},\vec{\pm \vec{K}}}$. 
The coupling between $\ket{\pm \vec{K},\text{TMD}}$ can be then captured by a two-step hopping, $T_{\vec{\Gamma,\vec{\pm K}}}T^\dagger_{\vec{\Gamma,\vec{\mp K}}}$, mediated by MPX$_3$ as shown in Fig.\ref{fig:MoireOffD}(c). 
We can apply second order perturbation theory to obtain $\Delta_\text{int}=\frac{1}{2}T^\dagger_{\vec{\Gamma,\vec{K}}} T_{\vec{\Gamma,\vec{K}'}}\left(\frac{1}{E_\vec{K}-E_\vec{\Gamma,\text{MPX$_3$}}}+\frac{1}{E_{\vec{K}'}-E_\vec{\Gamma,\text{MPX$_3$}}}\right)$ (see appendix). 
This analysis agrees well with our first-principles calculations, in which $\Delta_\text{int}$ is indeed inversely proportional to the energy difference between $\ket{\vec{K},\text{TMD}}$ and $\ket{\vec{\Gamma},\text{MPX}_3}$.
In the MoTe$_2$/MnPSe$_3$ heterostructure, our fitting shows that the relevant $\ket{\vec{\Gamma},\text{MnPSe}_3}$ band is about $0.66$ eV below the $\ket{\pm \vec{K},\text{MoTe}_2}$ bands at zero displacement field, consistent with the band structure in Fig.\ref{fig:MoireDia}(b). \\
\indent The global phase of the valley pseudospin texture is locked to the N\'eel vector of MPX$_3$. 
As the N\'{e}el vector is rotated from the $+x$ axis (which is the direction of the N\'eel vector for prior calculations and modeling) counter-clockwise by $\theta$, its effect to $\Delta_\text{int}$ can be analyzed from the spin dependence of the hopping channel, i.e., $T_{\vec{\Gamma},\vec{K}'}\propto \braket{\chi}{\downarrow}$, where $\ket{\chi}$ is the spinor component of the relevant magnetic bands in MPX$_3$, and its direction is mostly collinear with the magnetic order. 
When the N\'{e}el vector is rotated by $\theta$, the spinor inner product acquires a phase: $\braket{U^{-1}(\theta)\chi}{\downarrow}=\braket{\chi}{U(\theta)\downarrow}=\braket{\chi}{\downarrow}e^{i\theta/2}$. Combined with the phase factor from the other hopping channel $T_{\vec{\Gamma},\vec{K}}^\dagger$, one finds that $\Delta_\text{int}\rightarrow \Delta_\text{int}e^{i\theta}$. 
Thus, the phase factor of the valley mixing can be manipulated by the direction of the N\'eel vector of MPX$_3$. 
In particular, MnPSe$_3$ is experimentally probed to be a candidate material of XY model where the direction of its N\'eel vector can be tuned by strain\cite{ni2021imaging,jeevanandam1999magnetism}, which makes the phase tuning of $\Delta_\text{int}$ experimentally feasible.\\
\indent As every term in equation (\ref{eqn:Ham}) has been determined from first principle, we calculate and obtain the moir\'e bands using the plane-wave basis\cite{wu2018hubbard} (see appendix). In Fig.\ref{fig:MoireBandstr}(a), we compare the moir\'e band structures with and without intervalley coupling. Without intervalley coupling, the moir\'e bands from $\vec{K}$ and $\vec{K}'$ valleys cross each other in between $\vec{\kappa}'$ and $\vec{\kappa}'_-$. With intervalley coupling, the crossings are avoided and a local band gap of about $1$ meV is formed. Near the avoided crossing, the valley components of the moir\'e bands transition from $\vec{\kappa}$($\vec{\kappa}'$) component to $\vec{\kappa}'$($\vec{\kappa}$) component. \\
\indent Without external magnetic field, however, the top two moir\'{e} bands are degenerate at $\vec{\gamma}$ due to a combined symmetry of time reversal and mirror (along three A-B-C paths generated by $\hat{\mathcal{C}}_3$) in the moir\'{e} superlattices. To lift the degeneracy at $\vec{\gamma}$, we apply an out-of-plane magnetic field to create a Zeeman splitting between the two valleys, i.e.

\begin{align}\label{eqn:Ham+Eg}
    H_\text{eff}(\vec{r})=\begin{bmatrix}
     H_{\vec{k}-\vec{\kappa}}+V_\text{d}(\vec{r})+E_g\ & \Delta_\text{int}(\vec{r}) \\
    \Delta_\text{int}^\dagger(\vec{r}) & H_{\vec{k}+\vec{\kappa}}+V_\text{d}(\vec{r})-E_g\
    \end{bmatrix}.
\end{align}

As $\Delta_{\text{int}}$ and $E_g$ are fully tunable by electric and magnetic field, respectively, we explore the topological phases of the bands as these two parameters are tuned. 
We first check the band structure at $\Delta=-0.7$ meV and $E_g=3.45$ meV. 
Under this parameter setting, the topmost moir\'e band becomes fully isolated as in Fig.\ref{fig:MoireBandstr}(b). 
We then calculate the Berry curvature and integrated it over the moir\'e Brillouin zone\cite{xiao2010berry}. 
We find that the first band has a Chern number of $1$, while the second band carries a Chern number of $-1$. 
The Berry curvature is maximized around $\vec{\kappa}'$ where the band inversion occurs as in Fig.\ref{fig:MoireBandstr}(c).
When the Fermi level is tuned between the two bands, this heterostructure becomes a Chern insulator with a band gap of about $0.1$ meV.\\

\begin{figure}[H]
\includegraphics[width=\columnwidth]{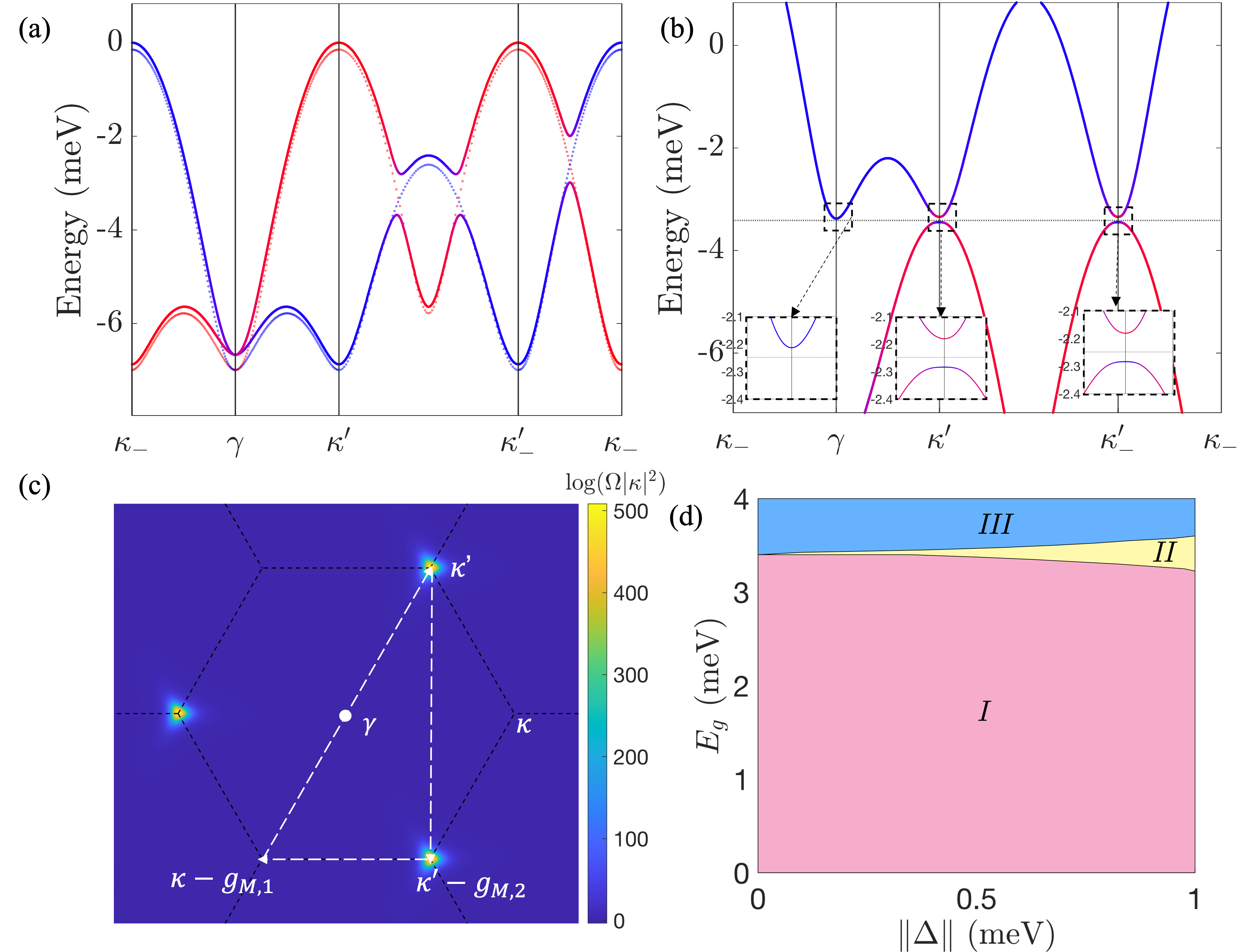}
\caption{\label{fig:MoireBandstr}(a) Moir\'{e} bandstructure of MoTe$_2$/MnPSe$_3$. Blue(red) represents $\vec{\kappa}$($\vec{\kappa}'$) band. Solid(dahsed) lines are bands with(without) intervalley coupling computed from equation(\ref{eqn:OffDiaPot}). The topmost bands are degenerate at $\vec{\gamma}$.  (b) Same as (a) but with external displacement field and intervalley Zeeman splitting. When the Fermi level is tuned between the first and second band, the system becomes a Chern insulator. Insets show zoomed-in gaps. (c) Normalized Berry curvature over the moir\'e Brillouin zone decorated with the k-path for (a) and (b): $\vec{\kappa}-\vec{g}_{M,1}(\vec{\kappa}_-)\rightarrow \vec{\gamma} \rightarrow \vec{\kappa}'\rightarrow \vec{\kappa}'-\vec{g}_{M,2}(\vec{\kappa}'_-)\rightarrow\vec{\kappa}-\vec{g}_{M,1}(\vec{\kappa}_-)$ (d) Phase diagram in the space spanned by $E_g$ and $\|\Delta\|$. Region \textit{I} (in red) denotes that the first two moir\'e bands are staggered with no global gap, and the Chern number of the first (second) band is $1$ ($-1$). Region \textit{II} (in yellow) denotes a quantum anomalous Hall insulator. Region \textit{III} (in blue) denotes a band insulator.} 
\end{figure}

\indent As mentioned previously, an out-of-plane magnetic field can break the combined time reversal and mirror symmetry. 
However, to reach the desired Zeeman splitting of about $3.5$ meV for the Chern insulator phase, a magnetic field of $~10$ T is required, even considering the large $g$-factor of the MoTe$_2$ valleys\cite{aivazian2015magnetic,deilmann2020ab,robert2021measurement}. 
This challenge can be resolved in two ways. 
First, we can cap MoTe$_2$/MnPSe$_3$ using another easy axis vdW magnet, e.g. CrBr$_3$, which can creates valley splittings as large as $10$ meV\cite{zhang2020abundant}. 
Second, an out-of-plane magnetic field can cant the magnetic moments in MnPSe$_3$ along the $z-$direction, and the proximity effect would induce a large effective Zeeman splitting in the MoTe$_2$ valleys.\\
\indent The highly tunable pseudospin textures allow us to control the topological phases of the Kekul\'e moir\'e superlattices as a function $\Delta$ and $E_g$ (Fig.\ref{fig:MoireBandstr}(d)). 
In region \textit{I}, under small $E_g$, the top two moir\'e bands are staggered and a global gap is absent, but the Chern number can be readily computed to be $1$($-1$) for the first (second) band. 
With increasing $E_g$, a global gap appears between the first and second band in region \textit{II}.
Since no band crossing happens yet, these bands remain topologically non-trivial. 
In this region, for one hole doping per moir\'e supercell, the system is in the Chern insulator phase as exemplified by Fig.\ref{fig:MoireBandstr}(b). 
At even higher $E_g$, the band gap closes, and after which trivial insulator phase kicks in at region \textit{III}.
As $\Delta$ increases, the allowed range of $E_g$ to realize the Chern insulator phases and the topological gap of the Chern insulator both increase.
As the topological gap in the Kekul\'e moire superlattice arises from intervalley coupling mediated by the magnetic substrate, such gap can survive even at temperatures much larger than the gap size, as long as the magnetic order stays.\\
\section{Summary}\label{sec3}
We have introduced a new type of moir\'e materials, where the lattices of the two constituent layers are related by a Kekul\'e reconstruction. 
Using MoTe$_2$/MnPSe$_3$ as an example, we have shown that the magnetism in MnPSe$_3$ enables the coupling between the Kramers' valley-spin states in MoTe$_2$. 
Lastly, we have constructed the effective moir\'e Hamiltonian, and show that the fascinating valley-pseudospin texture and non-trivial band topology can be tuned on-demand. 
The ability to fully control the Kramers’ valley-spin pairs could provide a new basis to enrich the moir\'{e} playground for quantum information science and materials research. \\
 
\section{Appendix}
\subsection{First Principle Calculations}
\indent  In the first-principles calculations of the $12\times 12$ commensurate structures based on local stacking geometry in the moir\'e supercell, we use the lattice constant $\sqrt{3}a_\text{MoTe$_2$}=6.11$ {\AA} based on $a_\text{MoTe$_2$}=3.53$ {\AA} obtained from relaxation. 
We use Perdew-Burke-Ernzehof (PBE)\cite{perdew1996generalized} for the exchange-correlation functional as implemented in Vienna \textit{ab initio} simulation package (VASP)\cite{kresse1996efficiency}. 
We employ projector augmented wave (PAW) pseudopotentials\cite{kresse1999ultrasoft} with a plane-wave energy cutoff for the wavefunctions of $500$ eV. 
The $k-$mesh is set to $9\times9$ and centered around $\vec{\Gamma}$. The internal atomic coordinates are fully relaxed until the force on every atom is small than $0.02$ eV{/\AA}. 
Van der Waals interactions are included within the DFT-D2 framework\cite{grimme2006semiempirical}. 
We set the vacuum thickness to be $18$ {\AA} to avoid interactions between repeated images. 
For the self-consistent calculation, a fully relativistic calculation is performed to include spin-orbit coupling. 
To accurately capture the magnetism of MnPSe$_3$, we apply the GGA plus on-site Hubbard $U$ method (GGA+U)\cite{liechtenstein1995density} where $U$ is set to $4$ eV for the Mn ions similar to previous calculations\cite{wang2006oxidation}. 
We note that the DFT calculations usually underestimate the quasiparticle bandgap, compared to the GW calculations which are extremely high in computational cost in this heterostructure.
However, the physics demonstrated in our work does not rely on the absolute value of the band gap. The important intervalley coupling arising from the magnetic substrate would not be affected, because such couplings act on degenerate valleys with the same quasiparticle self-energy correction.

\subsection{Moir\'{e} Lattice Vector}
In Kekul\'e moir\'e superlattices, the lattice constants of the larger (smaller) lattice are denoted as $\{\vec{a}_{\text{L},1},\vec{a}_{\text{L},2}\}$ ($\{\vec{a}_{\text{S},1},\vec{a}_{\text{S},2}\}$), and their directions can be referred to the inset of Fig. \ref{fig:LatticeStruct}(a). The superlattice formed by stacking two hexagonal lattice is also hexagonal. However, the moir\'{e} superlattice constant extracted based on the reconstructed Kekul\'e cell using the common continuum formula, i.e., $\displaystyle L_M=\frac{a_\text{L}}{\sqrt{\delta^2+\theta^2}}$, is off from the true value by $\sqrt{3}$. 
Here, $\displaystyle \delta\equiv\frac{\vert a_\text{L}-\sqrt{3}a_\text{S}\vert}{a_\text{L}}$ and $\theta$ is measured from the $30^\circ$ angle. This discrepancy can be solved from carefully analyzing the displacement vector. When we start from a lattice point, and travel by $u\vec{a}_{\text{L},1}+v\vec{a}_{\text{L},2}$, this point is off from the lattice point at $\displaystyle u\left(\vec{a}_{\text{S},1}+\vec{a}_{\text{S},2}\right)+v\left(2\vec{a}_{\text{L},1}-\vec{a}_{\text{S},2}\right)$ by the displacement vector, $\vec{d}(u,v)$. Explicitly, the displacement vector reads

\begin{align*}
    \vec{d}(u,v)&=\delta \left(u\vec{a}_{\text{L},1}+v\vec{a}_{\text{L},2}\right).
\end{align*}
When we transverse to a certain $(u,v)$, if $\vec{d}(u,v)$ is equal to some integer combination of $\{\vec{a}_{\text{L},1},\vec{a}_{\text{L},2}\}$, then the lattice points of the larger lattice and the smaller lattice at $u\vec{a}_{\text{L},1}+v\vec{a}_{\text{L},2}$ should overlap. This is exactly a periodicity at the wavelength of $u\vec{a}_{\text{L},1}+v\vec{a}_{\text{L},2}$. The true moir\'e periodicity should happen with $\vec{d}(u,v)$ equal to exactly one unit cell lattice constant. If we try $\vec{d}(u,v)=\vec{a}_{\text{L},1}$, we get $\displaystyle u=\frac{1}{\delta}$, and $\displaystyle L_M=\|u\hat{a}_{\text{L},1}\|=\frac{a_\text{L}}{\delta}$, which is still the value that is off by a factor of $\sqrt{3}$. The remedy comes from the fact that the smallest unit cell lattice constant is actually $\vec{a}_\text{S}$. When $\vec{d}(u,v)=\vec{a}_{\text{S},1}$, $\displaystyle u=v=\frac{a_\text{S}}{\sqrt{3}\delta }$, and $\displaystyle L_M=\left\|u\hat{\vec{a}}_{\text{L},1}+v\hat{\vec{a}}_{\text{L},2}\right\|=\frac{a_\text{S}}{\delta}$. Indeed, to construct the smallest moir\'{e} period, the displacement vector is equal to one primitive lattice vector of TMD, which in turn ensures that the primitive lattice vectors of the moir\'{e} lattice are parallel to those of the TMD. Combining with a twist angle, the true moir\'e period is given by
\begin{equation}
    L_M=\frac{a_\text{S}}{\sqrt{\delta^2+\theta^2}}.
\end{equation}\\

\indent At small lattice mismatch/twist angle, the continuum model is usually constructed with locally commensurate structure, and each local structure can be constructed by displacing one unit cell relative to another. The conventional real space moir\'{e} unit cell is then fully captured when the displacement vector scans the unit cell of the commensurate structure. In the case of Kekul\'e moir\'{e}, the displacement vector only needs to scan the unit cell of the TMD, which is one third of the area of the locally commensurate unit cell.
\subsection{Microscopic Theory of Effective Intervalley Coupling}
In the main text, we used first-principles calculations to obtain the off-diagonal parts of the moir\'e Hamiltonian.
Alternatively, The interlayer hopping matrix can be computed from the two-center approximation\cite{bistritzer2011moire,wang2017interlayer,wu2019topological}. 
We first write down the sublattice resolved plane wave basis for TMD centered around $\pm \vec{K}$ and MPX$_3$ centered around $\vec{\Gamma}$ as
\begin{align}
    \ket{\psi_{n,\eta \vec{K}+\vec{k}}^{\alpha, \text{S}}(\vec{r})}&=\frac{1}{\sqrt{N}}\sum_\vec{R}e^{i(\eta\vec{K}+\vec{k})\cdot(\vec{R}+\vec{\tau}_\alpha)}\ket{n(\vec{r}-\vec{R}-\vec{\tau}_\alpha)}\otimes\ket{\sigma_\eta},\\
    \ket{\psi_{m,\vec{\Gamma}+\vec{k}'}^{\beta, \text{L}}(\vec{r})}&=\frac{1}{\sqrt{N'}}\sum_\vec{R'}e^{i(\vec{\Gamma}+\vec{k}')\cdot(\vec{R}'+\vec{\tau}_\beta)}\ket{m(\vec{r}-\vec{R}'-\vec{\tau}_\beta)}\otimes\ket{\sigma_\beta}.
\end{align}
S (L), $\alpha$ ($\beta$), $\vec{\tau}_\alpha$ ($\vec{\tau}_\beta$), $n$ ($m$), $N$ ($N'$), $\vec{R}$ ($\vec{R}'$), $\vec{k}$ ($\vec{k}'$) denote the smaller (larger) lattice, sublattice indices, sublattice coordinates, band indices, number of unit cells, lattice vectors, crystal momenta of TMD(MPX$_3$) respectively. 
$\eta=\pm$ denotes the valley degree of freedom. 
The wavefunctions can be decomposed into two parts: the local orbital part $\ket{n}$ ($\ket{m}$) and the spinor part $\ket{\sigma_\eta}$ ($\ket{\sigma_\beta}$). 
Notably, for TMD the spinor part depends on the valley index $\eta$, while for MPX$_3$ the spinor part depends on the sublattice index.\\
The hopping channel can be then expressed as
\begin{align*}   T^{\alpha,\beta}_{\vec{\Gamma}\rightarrow\eta\vec{K}}&\equiv \braket{\psi_{n,\eta \vec{K}+\vec{k}}^{\alpha, \text{S}}(\vec{r})}{\psi_{m,\vec{\Gamma}+\vec{k}'}^{\beta, \text{L}}(\vec{r})}\\
&=\frac{1}{\sqrt{NN'}}\sum_{\vec{R},\vec{R}'}e^{-i(\eta\vec{K}+\vec{k})\cdot(\vec{R}+\vec{\tau}_\alpha)+i\vec{k}'\cdot(\vec{R}'+\vec{\tau}_\beta)}\\
&\times \braket{n(\vec{r}-\vec{R}-\vec{\tau}_\alpha)}{m(\vec{r}-\vec{R}'-\vec{\tau}_\beta)}\times \braket{\sigma_\eta}{\sigma_\beta}.
\end{align*}
The two-center approximation takes $\braket{n(\vec{r}-\vec{R}-\vec{\tau}_\alpha)}{m(\vec{r}-\vec{R}'-\vec{\tau}_\beta)}$ as $t_{nm}(\vec{R}+\vec{\tau}_\alpha-\vec{R}'-\vec{\tau}_\beta)$. Then after Fourier transform, we arrive at the final expression:
\begin{align*}  T^{\alpha,\beta}_{\vec{\Gamma}\rightarrow\eta\vec{K}}&=\frac{1}{\sqrt{\Omega\Omega'}}\sum_{\vec{G}_\text{S},\vec{G}_\text{L}}\delta_{\eta\vec{K}+\vec{k}+\vec{G}_\text{S},\vec{k}'+\vec{G}_\text{L}}t_{nm}(\eta\vec{K}+\vec{k}+\vec{G}_\text{S})\\
&\times e^{i\vec{G}_\text{S}\cdot \vec{\tau}_\alpha}e^{-i\vec{G}_\text{L}\cdot \vec{\tau}_\beta}\braket{\sigma_\eta}{\sigma_\beta}.
\end{align*}
Here, $\Omega$ ($\Omega'$) is the unit cell area of TMD (MPX$_3$); $\vec{G}_\text{S}$ ($\vec{G}_\text{L}$) is the reciprocal lattices of TMD (MPX$_3$); $t_{nm}(\vec{q})$ is the Fourier transformed amplitude of $t_{nm}(\vec{r})$. The full hopping channel matrix is obtained by identifying the sublattice indices $\alpha$, $\beta$ as matrix indices. For example, in MoTe$_2$/MnPSe$_3$, $T_{\Gamma\rightarrow\eta\vec{K}}$ will be a $3\times 2$ matrix as it considers a hopping from two magnetic Mn ions to three Mo ions. \\
The real space Hamiltonian in this formalism reads:
\begin{align*}
    H(\vec{r})=\begin{bmatrix}
        H_\vec{\Gamma}(\vec{r})&T^\dagger_{\Gamma\rightarrow\vec{K}}&T^\dagger_{\Gamma\rightarrow\vec{K}'}\\
        T_{\Gamma\rightarrow\vec{K}}&H_\vec{K}(\vec{r})&0\\
        T_{\Gamma\rightarrow\vec{K}'} & 0 & H_\vec{K'}(\vec{r})
    \end{bmatrix}.
\end{align*}
Note that each matrix element is by itself a matrix. In  MoTe$_2$/MnPSe$_3$ Kekul\'e moir\'e, $H_\vec{\Gamma}$ is a $2\times 2$ matrix, while $H_\vec{K}(\vec{r})$ is a $3\times3$ matrix. These diagonal matrix elements denote the standalone monolayer Hamiltonians. The intervalley coupling effect can then be obtained using a downfolding technique:
\begin{align*}
    H_\text{eff}(\vec{r})&=\begin{bmatrix}
        H_\vec{K}+C_{\vec{K}\vec{K}}\times \frac{1}{2}T_{\Gamma\rightarrow\vec{K}}T_{\Gamma\rightarrow\vec{K}}^\dagger& C_{\vec{K}\vec{K}'}\times \frac{1}{2}T_{\Gamma\rightarrow\vec{K}}T_{\Gamma\rightarrow\vec{K}'}^\dagger\\
        C_{\vec{K}'\vec{K}}\times \frac{1}{2}T_{\Gamma\rightarrow\vec{K}}T_{\Gamma\rightarrow\vec{K}'}^\dagger&H_{\vec{K}'}+C_{\vec{K}'\vec{K}'}\times \frac{1}{2}T_{\Gamma\rightarrow\vec{K}'}T_{\Gamma\rightarrow\vec{K'}}^\dagger
    \end{bmatrix},
\end{align*}
\begin{align*}
    C_{\eta\vec{K},\eta'\vec{K}}&=\left(\frac{1}{H_{\eta\vec{K}}-H_\vec{\Gamma}}+\frac{1}{H_{\eta'\vec{K}}-H_\vec{\Gamma}}\right).
\end{align*}
We recognize that the two-step hopping would provide both the diagonal and off-diagonal moir\'e potential ($V_\text{d}$ and $\Delta_\text{int}$). However, there's a caveat to it as this formalism treats the two layers as rigid bodies, i.e. there's no corrugation effects. As a result, it would predict that $V_\text{d}$ and $\Delta_\text{int}$ are on the same order of magnitude, while DFT calculation (which includes corrugation effects) predicts $V_\text{d}$ to be about one order of magnitude larger than $\Delta_\text{int}$. 
\subsection{Intervalley Coupling Selection Rule}
\indent Since $T_{\vec{\Gamma},\vec{\pm \vec{K}}}$ depend on the magnetic order in MPX$_3$ as well as the local stackings, under some configurations $\Delta_\text{int}$ vanishes by symmetry. 
For example, as shown in the main text, the out-of-plane magnetic orders are not capable of connecting $\ket{\pm\vec{K}}$ because the spinor product in one of $T_{\vec{\Gamma},\vec{\pm \vec{K}}}$ becomes zero.
We then notice that to lift the degeneracy of the $\ket{\vec{K}}-\ket{\vec{K}'}$ Kramers' pair, the time reversal symmetry, $\hat{\mathcal{T}}$, needs to be broken, where this condition is indeed fulfilled in all AFM orders. However, the time reversal with a sublattice translation is still a symmetry for some of the AFM orders. Notably, the AFM stripy phase and zigzag phase are invariant under $\hat{\mathcal{T}}\hat{\tau}_{\vec{a}_{\text{MPX}_3,2}/2}$, where $\vec{a}_{\text{MPX}_3,2}$ is the lattice vector in the $\left(\frac{1}{2},\frac{\sqrt{3}}{2}\right)$ direction. In the Bloch wave representation, $\hat{\tau}_{\vec{a}_{\text{MPX}_3,2}/2}$ becomes a phase factor, thus $\hat{\mathcal{T}}\hat{\tau}_{\vec{a}_{\text{MPX}_3,2}/2}\ket{\vec{K}}$ still maps to $\ket{\vec{K}'}$ up to a phase factor. In other words, $\ket{\vec{K}}$ and $\ket{\vec{K}'}$ still form an effective Kramers' pair, and the degeneracy is intact. For N\'{e}el AFM order, however, no such symmetry as $\hat{\mathcal{T}}\hat{\tau}_{\vec{a}_{\text{MPX}_3,2}/2}$ is present. As a result, we conclude that $\Delta_\text{int}$ can only occur in in-plane N\'eel AFM and FM orders.\\

\subsection{Plane Wave Basis Method}
As the moir\'e Hamiltonian is translational invariant under $\vec{a}_M$, we can diagonalize the moir\'e Hamiltonian using the moir\'e wavevector basis, $\{\ket{\vec{\vec{k}-\eta \kappa +\vec{g}_M}}\}$, where $\eta=\pm 1$ denotes the two different valleys and $\vec{g}_M$ is a moir\'e reciprocal lattice vector \cite{wu2018hubbard}. 
For the diagonal terms, the non-zero matrix elements are the ones from the same valley
\begin{align*}
    \bra{\vec{k}-\vec{\kappa} +\vec{g}_M}H_{\vec{k}-\vec{\kappa}}\ket{\vec{k}-\vec{\kappa} +\vec{g}'_M}&=-\delta_{\vec{g}_M,\vec{g}'_M}\frac{\hbar^2(\vec{k}-\vec{\kappa})^2}{2m^*},\\
    \bra{\vec{k}+\vec{\kappa} +\vec{g}_M}H_{\vec{k}+\vec{\kappa}}\ket{\vec{k}+\vec{\kappa} +\vec{g}'_M}&=-\delta_{\vec{g}_M,\vec{g}'_M}\frac{\hbar^2(\vec{k}+\vec{\kappa})^2}{2m^*},\\
    \bra{\vec{k}-\eta\vec{\kappa} +\vec{g}_M}V_\text{d}\ket{\vec{k}-\eta\vec{\kappa} +\vec{g}'_M}&=V_\text{d}(\vec{g}_M-\vec{g}'_M).
\end{align*}
Here, $V_\text{d}(\vec{g}_M-\vec{g}'_M)$ is the Fourier transformed $V_\text{d}(\vec{r})$. Similarly, for the off-diagonal terms, the matrix element is only non-zero between states from different valleys:
\begin{align*}
    \bra{\vec{k}-\vec{\kappa} +\vec{g}_M}\Delta_\text{int}\ket{\vec{k}+\vec{\kappa} +\vec{g}'_M}=\Delta_\text{int}(\vec{g}_M-\vec{g}'_M),\\
    \bra{\vec{k}+\vec{\kappa} +\vec{g}_M}\Delta^\dagger_\text{int}\ket{\vec{k}-\vec{\kappa} +\vec{g}'_M}=\Delta_\text{int}^*(\vec{g}_M-\vec{g}'_M).
\end{align*}
Again, $\Delta_\text{int}(\vec{g}_M-\vec{g}'_M)$ is nothing but the Fourier transformed $\Delta_\text{int}(\vec{r})$. \\
\indent With knowing these matrix elements, one can write down the moir\'e Hamiltonian and the basis set in the $\vec{k}-$space with any number of Brillouin zones. 
In this work, the calculation is carried out with $7\times 7$ moir\'e Brillouin zone to reduce the finite size errors for the first few moir\'e bands.
\section*{Acknowledgments}
The first-principles calculations of two-dimensional magnetic materials was supported by the University of Washington Molecular Engineering Materials Center (DMR-1719797). The theoretical study of Kekul\'e moir\'e Hamiltonian is supported by the Department of Energy BES QIS program on `Van der Waals Reprogrammable Quantum Simulator' under award number DE-SC0022277. This work was facilitated through the use of advanced computational, storage, and networking infrastructure provided by the Hyak supercomputer system and funded by the University of Washington Molecular Engineering Materials Center at the University of Washington.

\printbibliography

\end{document}